# Vers une interface pour l'enrichissement des requêtes en arabe dans un système de recherche d'information

Mohammed El Amine Abderrahim

Université Abou Bekr Belkaid Tlemcen, Algérie
Faculté des sciences de l'ingénieur
Département d'informatique
BP 230 chetouane
medamineabd@yahoo.fr

**Résumé.** Dans le cadre de la Recherche d'Information (RI) pour les textes en langue arabe, nous proposons dans le présent article la réalisation d'une interface utilisateur qui emploi un analyseur morphologique pour récupérer les formes de base des mots arabe présents dans la requête de l'utilisateur, pour ensuite faire appel à WordNet Arabe pour enclencher le processus d'expansion. La requête ainsi étendue est envoyée à Google. Par ce modeste travail, nous espérons apporter un premier pas vers l'utilisation des méthodes linguistiques et des ressources lexicales pour l'enrichissement de requête dans un système de RI arabe.

**Mots Clés :** Recherche d'Information Arabe, TALN Arabe, expansion de la requête, Wordnet Arabe.

**Abstract.** This presentation focuses on the automatic expansion of Arabic request using morphological analyzer and Arabic Wordnet. The expanded request is sent to Google.

## 1. Introduction

Un Système de Recherche d'Information (SRI) repose sur les trois fonctions suivantes : stocker, organiser (indexer) et rechercher des données (en réponse à des requêtes utilisateurs). Il doit faire appel à trois types de connaissances:

- les connaissances sur les documents : ils regroupent les informations sur le contenu et le contenant ;
- les connaissances sur les utilisateurs ;
- et les connaissances sur le domaine d'application : ils permettent d'organiser les différents termes utilisés, on retrouve par exemple les dictionnaires, les thesaurus…

Dans le cadre de la recherche d'information pour les textes en langue arabe, la récupération de mots clé est jugée insuffisante, car les termes utilisés dans la requête peuvent présenter par rapport aux documents de la base, des différences sur plusieurs plans, par exemple :
- des variations morphologiques comme dans « مدرسة » et « مدرستان », « خيل » et « خيول » ;
- des variations lexicales (on utilise pour le même sens des mots différents) comme dans le cas dans « خيل » et « فرس » ;
- des variations sémantiques comme dans le cas de « مرادف : الـحـجر » et « الحجر: أنثى الخيـل » « الصــخــر ».

L'utilisation des ontologies pour l'enrichissement (expansion) de la requête utilisateur peut constituer une solution (parmi d'autres) pour résoudre le problème des variations sémantiques, en effet, les ontologies offrent des ressources sous la forme de relations sémantiques, ils permettent d'étendre le champ de recherche d'une requête, ce qui a pour conséquence l'amélioration des résultats de la recherche. Par ailleurs l'utilisation d'un analyseur morphologique peut suffire pour résoudre les deux premiers cas de variations (morphologiques et lexicales).

L'utilisation des ontologies dans un SRI peut être envisagé à plusieurs niveaux :
- avant d'être envoyée, la requête de l'utilisateur pourra être enrichie par les concepts jugés proches dans l'ontologie et ceci, par le biais de l'utilisation des relations comme la généralisation/spécialisation, la synonymie…
- L'indexation des documents se fait en utilisant les concepts de l'ontologie et non pas les mots clés.
- Le filtrage des documents selon un domaine particulier pour des profils d'utilisateur. [7], [8], [15], [13], [5], [19].

Il faut noter toutefois que la qualité des réponses obtenues par un SRI ne dépend pas seulement de la qualité du processus l'appariement requête/documents mais aussi de la requête formulée par l'utilisateur, d'où l'intérêt de la reformulation de la requête.
On distingue deux approches pour la reformulation d'une requête dans un SRI : directe et indirecte.
- Reformulation directe : elle consiste à ajouter de nouveaux termes à la requête initiale en s'appuyant sur des ressources lexicales comme les dictionnaires ou bien sur les liens de co-occurrences entre les termes.
- Reformulation indirecte : En tenant compte d'une liste de documents déjà jugés sélectionnés, la requête est modifiée. Ce processus est appelé réinjection de la pertinence « relevance feed-back ».

Dans cette optique, les travaux sur les SRI pour les textes en arabe ne sont pas nombreux à notre connaissance. Parmi ces travaux on trouve :
- le système de [10]. Ce dernier adopte la notion de schème comme base pour lemmatiser les mots et les substituer par leurs lemmes dans les opérations d'indexation et de recherche.
- Le système de [12]. Il propose d'assister l'utilisateur dans la formulation de sa requête par l'utilisation des modèles de n-gram. Pour les opérations d'indexation et de recherche, [12] utilise les services du moteur de recherche Google.

Cet article entre dans le cadre de l'assistance de l'utilisateur par l'amélioration de sa requête (reformulation directe) en utilisant un analyseur morphologique et une ressource lexicale (dans notre cas il s'agit de Wordnet Arabe). Dans ce qui suit nous allons décrire l'architecture et le mode de fonctionnement de notre interface pour l'enrichissement de la requête Arabe. Mais, commençant d'abord par une brève description de notre analyseur morphologique ainsi que la ressource utilisée wordnet arabe.

## 2. L'analyseur morphologique (AM)

L'entrée de l'AM est une requête écrite en arabe. Un premier traitement consiste à segmenter la requête en formes. Le séparateur blanc étant la marque des frontières des formes, ce traitement ne devra donc poser aucun problème. La figure suivante (voir figure 1) montre le principe de l'AM d'une requête qui se résume dans deux étapes :
1. consultation du lexique des mots outils (mots vides) ;
2. segmentation et analyse de la forme. L'opération de segmentation procède par l'accès aux différentes tables (clitiques et affixes) pour détecter la présence de proclitique, enclitique, préfixe et suffixe dans la forme. Le résultat étant un ensemble de cinq segments (proclitique, préfixe, radical, suffixe, enclitique). Par ailleurs, l'opération d'analyse effectue un accès au dictionnaire des formes simples pour vérifier l'existence du radical. S'il existe, l'analyseur lui associe l'ensemble de ses informations linguistiques (la base, racine…). [2].

Donc, a l'issue de l'AM de la requête, l'analyseur produit un ensemble d'informations (base, racine, catégorie grammaticale, ensemble de traits syntaxiques…) qui représente la solution morphologique hors-contexte calculée dans le modèle linguistique utilisé. Toutefois, ce qui nous intéresse dans le cadre de notre étude est évidement l'ensemble des formes de base composant la requête initiale de l'utilisateur. Le détaille de la procédure de l'analyse morphologique se trouve dans [2].

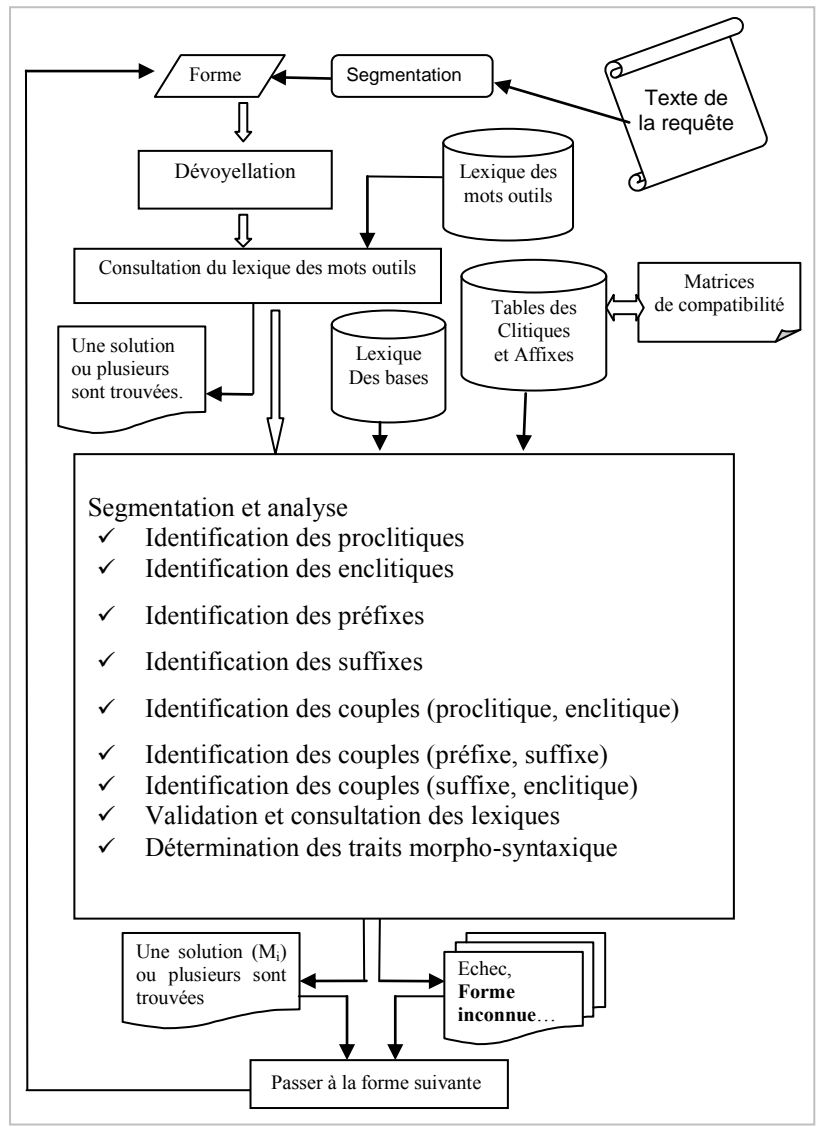

**Fig. 1.** Architecture générale de l'analyseur morphologique

## 3. Wordnet Arabe

Wordnet Arabe est une base de données lexicale librement disponible pour l'arabe standard. Cette base de données suit la conception et la méthodologie du Princeton Wordnet pour l'anglais et d'EuroWordnet pour les langues européennes. Sa structure est celle d'un thésaurus, il est organisé autour de la structure des synsets, c'est-à-dire des ensembles de synonymes et de pointeurs décrivant des relations vers d'autres synsets. Chaque mot peut appartenir à un ou plusieurs synsets, et à une ou plusieurs catégories du discours. Ces catégories sont au nombre de quatre : nom, verbe, adjectif et adverbe. Wordnet et donc un réseau lexical dont les synsets sont les nœuds et les relations entre synsets sont les arcs. Il faut noter toutefois que Wordnet Arabe est une des rares ressources pour la langue générale arabe disponible en ligne. Il compte actuellement[1] 11269 synsets et 23481 mots. [14], [16], [11], [17], [20], [21].

## 4. Architecture et fonctionnement de l'interface

Notre interface de recherche (voir la figure 2) se compose de deux modules importants : l'analyseur morphologique et le module de recherche des concepts à partir de Wordnet. Par ailleurs il utilise deux ressources de données:
- une base de données linguistique (contenant les différents lexiques ainsi que l'ensemble des clitiques et affixes propres à la langue arabe) utilisée par l'analyseur morphologique ; son contenu est détaillé dans [1], [2], [3].
- Wordnet Arabe.

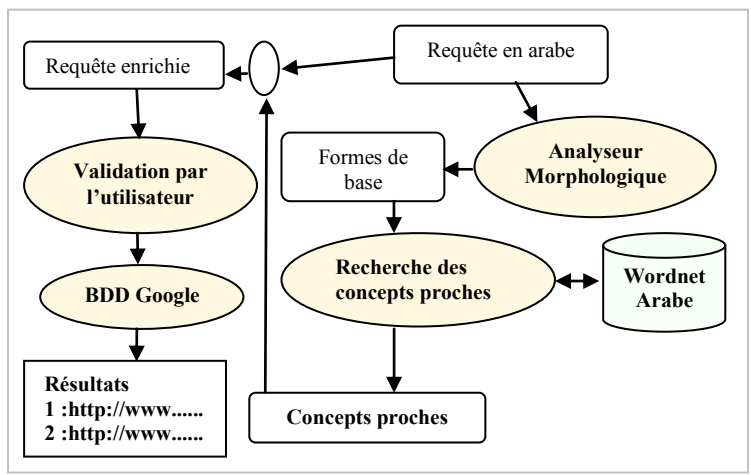

**Fig. 2.** Architecture de l'interface de recherche

---

[1]Mars 2009 ; voir la page http://www.lsi.upc.edu/~mbertran/

La construction de la liste des concepts les plus proche de la forme de base passe donc par Wordnet Arabe. Par exemple, à partir de la forme «درس » on construit la liste suivante (voir tableau 1).

**Tableau 1.** Concepts «درس » à partir de Wordnet Arabe

| N° | Concepts «درس» |
|----|----------------|
| 1  | طَالِب, بَاحِث, دَارِس |
| 2  | طَالِب, دَارِس, مُتَعَلِّم, تِلْمِيذ |
| 3  | دَرَسَ |
| 4  | أَخَذَ, دَرَسَ, قَرَأ, تَعَلَّم |
| 5  | بَحَثَ, دَرَسَ, رَاعَى |
| 6  | عَايَنَ, حَلَّلَ, دَرَسَ, فَحَصَ, نَاقَشَ |
| 7  | إِعْتَبَرَ, دَرَسَ, فَحَصَ, نَظَرَ فِي |
| 8  | دَرْس, دَوْرَة, دَوْرَة دِرَاسِيَّة, مُقَرَّر, مُقَرَّر تَعْلِيمِي, مُقَرَّر تَدْرِيسِي |
| 9  | دَرَّسَ |
| 10 | عَلَّمَ, دَرَّبَ, دَرَّسَ, هَذَّبَ, رَبَّى, ثَقَّفَ |
| 11 | دِرَاسَة, تَقْرِير, تَقْرِير كِتَابِي |
| 12 | بحث, بَحْث, دراسة |
| 13 | مَدْرَسَة |
| 14 | نَاقَشَ, تَشَاوَرَ, تَبَاحَثَ, تَدَارَسَ, تَدَاوَلَ, تَنَاقَشَ |
| 15 | تعليم, تَعْلِيم, تَدْرِيس, تَدْرِيس |

Après la saisie de la requête, le principe de fonctionnement de notre interface comprend deux alternatives :
- l'utilisateur ne veut pas utiliser le module d'enrichissement de la requête. C'est le cas le plus simple, il suffit donc d'envoyer la requête à Google. Pour cela, nous avons utilisé une API libre (disponible gratuitement en ligne) fourni par Google pour l'interrogation de sa base de données et la récupération des résultats.
- L'utilisateur veut enrichir sa requête (voir la figure 2). Dans ce cas le texte de la requête est envoyé à l'analyseur morphologique pour produire une liste de formes de base qui va servir à générer une liste des concepts proches en utilisant wordnet arabe. La liste ainsi générée ainsi que le texte de la requête initiale forment le texte de la requête enrichi. Cette dernière est envoyée à Google après sa validation par l'utilisateur.

## 5. Discussion

Les systèmes de recherche d'information classiques traitent la requête de manière à optimiser les temps de recherche et l'identification de documents selon des critères d'appariement entre les mots contenus dans les requêtes utilisateurs (et uniquement ceux-là) et ceux des documents.

Dans notre cas, nous nous intéressons à la formulation de requête : l'idée est donc d'exploiter le contenu de wordnet arabe et d'un analyseur morphologique pour reformuler et étendre des requêtes (par expansion) de manière à retrouver plus précisément les bons documents.
Un problème est donc traité, il s'agit de palier le problème des variations lexicales, autrement dit, notre interface interroge wordnet arabe pour récupérer des mots différant lexicalement, mais reliés à ceux de la requête initiale par des relations sémantiques, telles que la synonymie, la généralisation et la spécialisation.

Un problème que nous n'avons pas traité dans le cadre du présent travail s'est posé au niveau du choix du sens (synset) à prendre dans le cas de la polysémie dans les mots de la requête.

Le mode de fonctionnement de notre interface se résume ainsi : on commence la phase d'expansion de la requête en analysant les mots de la requête à l'aide de l'analyseur morphologique. wordnet arabe est ensuite interrogé pour récupérer une liste des termes reliés à la requête par des relations de synonymie, généralisation et spécialisation. L'ensemble des termes constitués par cette dernière liste ainsi que la liste des mots de la requête initiale forme ainsi la requête enrichie qui sera validée par l'utilisateur et envoyée au moteur de recherche Google.

L'évaluation de l'apport réel de l'enrichissement de la requête arabe est une tâche très délicate et demande par conséquent beaucoup d'investigations. Toutefois si nous allons se baser sur les études faites sur d'autres langues nous pouvons dire que l'apport des ontologie dans le domaine de la RI se caractérise par :
- réduction du silence dans les réponses aux requêtes utilisateurs ;
- réduction du nombre des réponses bruitées ;
- expression de la requête plus facilement (assistance dans la formulation de la requête);

Pour la confirmation de ces hypothèses, il nous reste maintenant les tâches suivantes :
- fixer le nombre de concepts qui doivent être choisi pour chaque relation utilisée (*synonymie, hypernymie, hyponymie…*),
- fixer le nombre de termes formant un concept d'extension résultant d'une relation sémantique,
- fixer les poids à affecter aux mots des concepts résultant de l'extension,
- l'étude de l'influence de l'utilisation des concepts composés de l'ontologie,
- l'étude de l'apport de chaque relation sémantique utilisée dans le processus d'enrichissement.

## 6. Conclusion

Le processus de recherche d'information se compose de trois parties : construire la requête, construire la réponse, évaluer la réponse. La qualité de la réponse dépend largement de la qualité de la requête construite, ainsi, une requête clairement formulée est beaucoup plus complexe que sa réponse. L'idée de cet article est d'exploiter une ressource lexicale et un analyseur morphologique pour reformuler (par expansion) la requête de l'utilisateur afin d'améliorer les résultats de la recherche. Pour tester cette approche nous avons utilisé le moteur de recherche Google avec wordnet arabe.
Une amélioration possible de notre travail consiste à palier le problème des variations morphologiques par l'exploitation des formes de base des mots de la requête produits par l'analyseur morphologique pour déduire les formes dérivées.

Beaucoup de tests et d'améliorations restent à faire et comme perspective nous sommes entrain de construire un corpus de texte arabe avec lequel nous pensons faire une évaluation objective de l'apport réel de cette approche dans un SRI pour les textes en langue arabe.

## Bibliographie